\documentclass[11pt]{article}

\usepackage{authblk}
\usepackage[utf8]{inputenc}
\usepackage[T1]{fontenc}
\usepackage{cite}
\usepackage{hyperref}
\usepackage{graphicx}
\usepackage{float}
\usepackage{amsmath}
\usepackage{booktabs}
\usepackage{array}
\usepackage{multirow}

\graphicspath{{figures/}}

\usepackage[margin=1in]{geometry}

\title{PrompTrend: Continuous Community-Driven Vulnerability Discovery and Assessment for Large Language Models}

\author[1]{Tarek Gasmi}
\author[2,3]{Ramzi Guesmi}
\author[4]{Mootez Aloui}
\author[5]{Jihene Bennaceur}

\affil[1]{University of Manouba, Tunisia}
\affil[2]{University of Jendouba, Tunisia}
\affil[3]{LETI Laboratory, University of Sfax, Tunisia}
\affil[4]{DataDoIt, Tunisia}
\affil[5]{South Mediterranean University, Tunisia}

\date{}

\begin{document}

\maketitle

\begin{abstract}
Static benchmarks fail to capture LLM vulnerabilities emerging through community experimentation in online forums. We present PrompTrend, a system that collects vulnerability data across platforms and evaluates them using multidimensional scoring, with an architecture designed for scalable monitoring. Cross-sectional analysis of 198 vulnerabilities collected from online communities over a five-month period (January-May 2025) and tested on nine commercial models reveals that advanced capabilities correlate with increased vulnerability in some architectures, psychological attacks significantly outperform technical exploits, and platform dynamics shape attack effectiveness with measurable model-specific patterns. The PrompTrend Vulnerability Assessment Framework achieves 78\% classification accuracy while revealing limited cross-model transferability, demonstrating that effective LLM security requires comprehensive socio-technical monitoring beyond traditional periodic assessment. Our findings challenge the assumption that capability advancement improves security and establish community-driven psychological manipulation as the dominant threat vector for current language models.
\end{abstract}


\section{Introduction}

The rapid deployment of Large Language Models (LLMs) across critical sectors—from healthcare and finance to education and public services—has created an unprecedented security challenge \cite{anthropic2022red,openai2023gpt4}. While academic researchers and corporate red teams work diligently to identify and mitigate vulnerabilities through controlled testing environments, a parallel universe of vulnerability discovery unfolds daily across Reddit threads, Discord servers, and Twitter conversations. In these digital spaces, thousands of users experiment with LLMs, share exploitation techniques, and collectively refine methods to bypass safety mechanisms—often weeks or months before these vulnerabilities appear in formal security assessments \cite{shen2024anything}. This disconnect between institutionalized security research and grassroots vulnerability discovery represents not merely a timing gap, but a fundamental blind spot in our approach to LLM safety.

The field of LLM security has evolved rapidly since the introduction of ChatGPT in late 2022, with researchers developing increasingly sophisticated frameworks for vulnerability assessment. Traditional approaches have focused on controlled red teaming exercises, where expert teams attempt to elicit harmful outputs through carefully crafted adversarial prompts \cite{anthropic2022red,perez2022red}. These efforts have produced valuable resources such as HarmBench \cite{mazeika2024harmbench}, AdvBench, and HELM Safety, which provide standardized datasets for evaluating model robustness. Concurrently, automated red teaming methods have emerged, employing techniques ranging from reinforcement learning \cite{perez2022red} to evolutionary algorithms \cite{samvelyan2024rainbow} to generate diverse attack vectors. Yet despite these advances, the security community continues to play catch-up with vulnerabilities that often originate not in research labs, but in online communities where users freely experiment with and discuss LLM limitations.

The limitations of current approaches become evident when examining recent vulnerability timelines. The ``DAN'' (Do Anything Now) jailbreak, for instance, was widely discussed and refined across multiple online platforms well before it was formally analyzed in academic or corporate settings \cite{shen2024anything}. Similarly, techniques for extracting training data, bypassing content filters through role-playing scenarios, and exploiting context window limitations often achieve viral status in user communities long before formal documentation \cite{wei2023jailbroken}. This pattern, illustrated in Figure 1, reveals a critical weakness in our security infrastructure: while we excel at analyzing known vulnerabilities in controlled settings, we lack systematic mechanisms for observing how these vulnerabilities actually emerge, evolve, and spread in the wild.

\begin{figure}[H]
\centering
\includegraphics[width=0.8\textwidth]{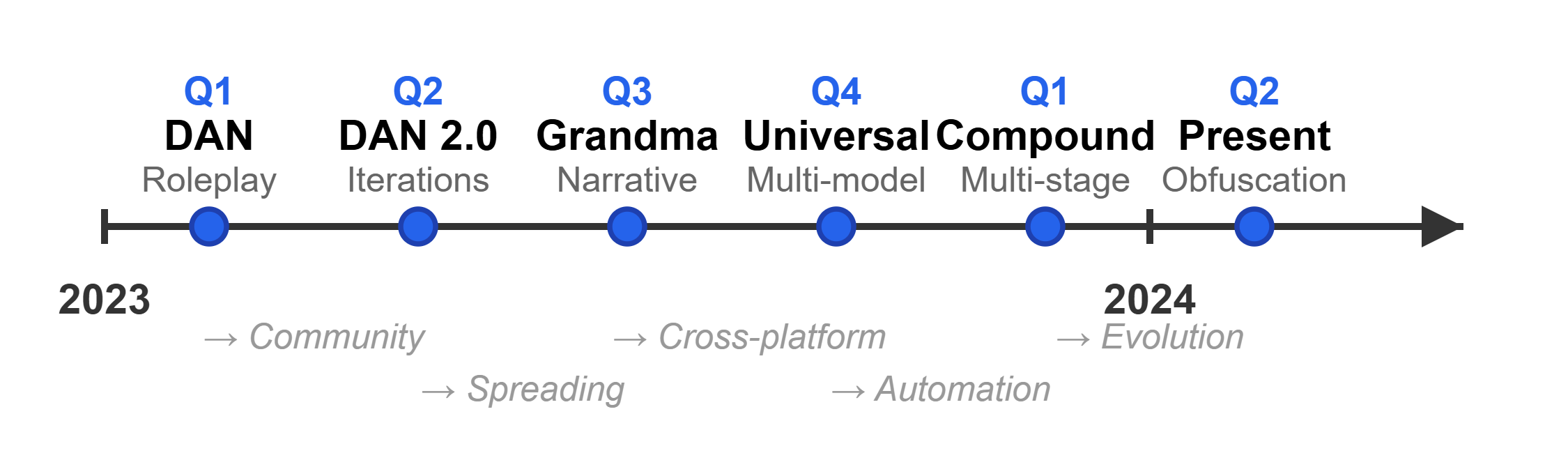}
\caption{Evolution Timeline of a Major Jailbreak Technique}
\label{fig:timeline}
\end{figure}

The research gap extends beyond mere timing delays. Current vulnerability assessment frameworks, exemplified by metrics like Attack Success Rate (ASR) and adaptations of the Common Vulnerability Scoring System (CVSS), fail to capture crucial dimensions of real-world threats \cite{first2023cvss}. These frameworks typically evaluate vulnerabilities in isolation, measuring technical characteristics while ignoring the social dynamics that determine whether a vulnerability will remain an academic curiosity or become a widely adopted threat. They provide snapshots of vulnerability effectiveness at specific points in time, but offer no insight into how these vulnerabilities persist or evolve as models are updated and defenses are implemented. Most critically, they cannot account for the collaborative refinement processes that occur when thousands of users iterate on and improve adversarial techniques through community feedback loops.

To address these critical gaps, we present PrompTrend, a comprehensive system for continuous monitoring and evaluation of LLM vulnerabilities as they emerge in online communities. As shown in Figure 2, PrompTrend deploys intelligent agents across multiple platforms to identify and collect adversarial prompts, with an architecture designed for real-time tracking, creating the first systematically collected dataset of in-the-wild vulnerability discoveries. Our research seeks to answer four fundamental questions:

\begin{itemize}
\item \textbf{RQ1:} How can real-time, community-driven intelligence contribute to the early detection of emerging vulnerabilities in LLMs?
\item \textbf{RQ2:} How can transformation-aware adversarial testing and multi-dimensional risk scoring improve the robustness and relevance of LLM vulnerability assessments?
\item \textbf{RQ3:} How can community-driven threat intelligence improve LLM robustness evaluation compared to static benchmarking approaches?
\item \textbf{RQ4:} What evaluation metrics best capture the practical effectiveness of vulnerability assessment frameworks in real-world deployment scenarios?
\end{itemize}

\begin{figure}[H]
\centering
\includegraphics[width=0.9\textwidth]{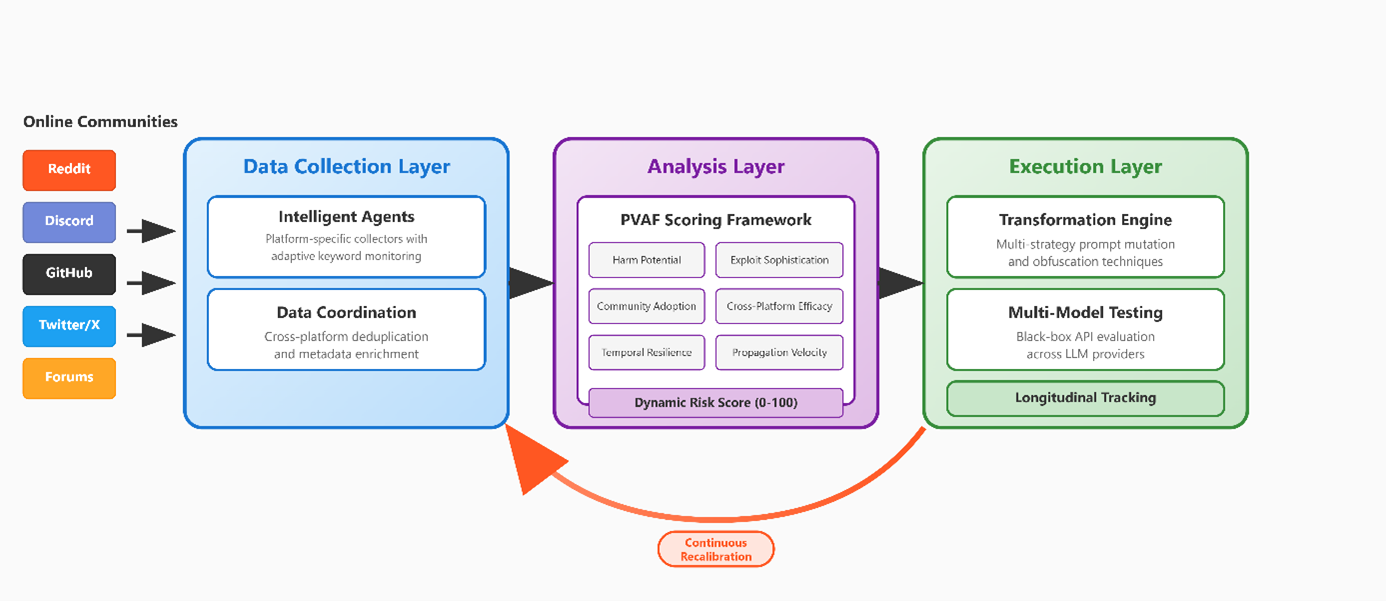}
\caption{Conceptual Overview of the PrompTrend System}
\label{fig:promptrend_overview}
\end{figure}

This paper makes four primary contributions to the field of AI safety research. First, we introduce a novel framework for real-time vulnerability discovery that bridges the gap between formal security research and grassroots exploration. Unlike traditional approaches that rely on periodic assessments or controlled testing, our system provides continuous visibility into the evolving threat landscape. Second, we present the PrompTrend Vulnerability Assessment Framework (PVAF), the first scoring system that incorporates both technical characteristics and social dynamics of vulnerability propagation. This framework recognizes that a vulnerability's real-world impact depends not only on its technical sophistication but also on factors like community adoption rates, cross-platform effectiveness, and temporal resilience. Third, PrompTrend establishes the first longitudinal dataset of community-discovered LLM vulnerabilities, enabling unprecedented analysis of how these threats evolve over time. This dataset captures not only the vulnerabilities themselves but also rich metadata about their discovery context, spread patterns, and effectiveness trajectories. Finally, our work represents a methodological advancement in cybersecurity research, demonstrating how observational studies of online communities can complement traditional security assessment approaches.

The remainder of this paper is organized as follows. Section 2 reviews related work in LLM security evaluation, social media intelligence, and vulnerability assessment frameworks. Section 3 details the PrompTrend system architecture, including our multi-agent design and the PVAF scoring framework. Section 4 describes our methodology for data collection and evaluation. Section 5 presents comprehensive results from our deployment, including empirical validation across multiple commercial LLMs. Section 6 discusses the implications of our findings and acknowledges limitations. Section 7 concludes with reflections on the future of community-integrated AI safety research.

\section{Background and Related Work}

This section examines the current state of LLM vulnerability assessment, highlighting the disconnect between formal security evaluation and real-world threat emergence. We analyze existing frameworks, their limitations, and the critical gap in community-driven threat intelligence that motivates our work.

\subsection{LLM Vulnerabilities and Attack Taxonomy}

The deployment of LLMs through commercial APIs has created novel security challenges fundamentally different from traditional software vulnerabilities \cite{yao2024survey}. Unlike systems with explicit access controls, LLMs process natural language probabilistically, making them susceptible to manipulation through carefully crafted linguistic inputs \cite{devore2011probability}. This vulnerability manifests primarily through two attack categories: jailbreaks and prompt injections.

Jailbreak attacks circumvent safety mechanisms through psychological manipulation, employing roleplay scenarios and narrative framing to exploit the model's tendency to maintain conversational consistency \cite{shen2024anything}. The ``Do Anything Now'' (DAN) jailbreak exemplifies this approach, creating fictional personas that operate outside normal constraints. These techniques have evolved rapidly through community experimentation, producing variants like the ``Grandma Hack'' and ``Simulation Mode'' that demonstrate increasing sophistication \cite{brown2024community}. Prompt injection attacks embed malicious instructions within legitimate queries, exploiting the fundamental characteristic that enables LLM functionality—their ability to interpret and execute complex contextual instructions \cite{greshake2023not}. These attacks succeed because LLMs process operational instructions and user content within the same input stream, making it difficult to distinguish between authorized commands and malicious manipulation.

At the implementation level, adversarial transformation methods disguise malicious content while preserving effectiveness. These include character-level obfuscation using Unicode substitution and zero-width spaces, encoding transformations through Base64 or hexadecimal conversion, and syntactic restructuring via paraphrasing or metaphorical language \cite{morris2023textattack}. Recent research reveals that successful attacks increasingly combine multiple techniques, with multi-turn manipulation showing 71\% higher success rates and non-English attacks demonstrating up to 195\% increased vulnerability \cite{amazon2024multilingual}. This evolution highlights the inadequacy of static, single-language evaluation approaches.

\subsection{Current Evaluation Frameworks and Their Limitations}

The rapid emergence of LLM vulnerabilities has prompted development of various evaluation frameworks, yet these approaches suffer from fundamental limitations when confronting the dynamic nature of real-world threats. Static benchmarking remains the dominant paradigm, exemplified by frameworks like HarmBench \cite{mazeika2024harmbench}, which provides standardized test cases across diverse harm categories. While valuable for systematic comparison, these benchmarks represent temporal snapshots that quickly become obsolete as new attack vectors emerge. The HELM Safety benchmark attempts broader coverage by aggregating multiple test suites, but still relies on predetermined cases that cannot capture evolving community-discovered vulnerabilities \cite{stanford2024helm}.

Automated red teaming methods have emerged to address the limitations of static datasets. Perez et al. \cite{perez2022red} pioneered using language models to attack other language models, enabling scalable vulnerability discovery. However, these approaches often suffer from mode collapse, generating limited attack diversity. The Rainbow Teaming framework introduced quality-diversity algorithms to explicitly optimize for both attack success and behavioral diversity \cite{samvelyan2024rainbow}, with RAINBOWPLUS further improving through multi-element archives and probabilistic fitness evaluation \cite{samvelyan2024rainbowplus}. Despite these advances, automated methods operate in controlled environments disconnected from the collaborative refinement processes that characterize real-world attack development.

Gradient-based optimization techniques like the Greedy Coordinate Gradient (GCG) method directly manipulate prompt components using model gradients \cite{zou2023universal}. While technically powerful, these approaches generate unnatural prompts with low perplexity, making them less representative of community-developed attacks and potentially easier to detect. More critically, all these methods focus on technical attack generation while ignoring the social dynamics that determine which vulnerabilities achieve widespread adoption.

Current vulnerability scoring predominantly relies on binary metrics like Attack Success Rate (ASR), which simply measures the percentage of successful attacks without capturing nuance \cite{carlini2017towards}. Recent attempts at sophisticated scoring include attention-based risk models \cite{pu2024feint} and gradient-based harmfulness detection \cite{xie2024gradsafe}, but these remain technically focused. The absence of frameworks incorporating social propagation, community adoption, and temporal persistence represents a critical gap, as these factors often predict real-world exploitation better than technical sophistication alone \cite{pfleeger2012leveraging}.

\subsection{The Community-Driven Threat Landscape}

The emergence of LLM vulnerabilities through online communities represents a paradigm shift from traditional security research. Unlike controlled laboratory discoveries, LLM attack techniques often originate through distributed experimentation across Reddit, Discord, Twitter, and GitHub \cite{reddit2024platform}. This collaborative development produces rapid innovation cycles where basic discoveries undergo community refinement, frequently yielding sophisticated attacks that individual researchers would not develop independently \cite{liu2018collaborative}.

Platform-specific dynamics significantly influence vulnerability evolution. Reddit communities serve as initial discovery grounds where users share experimental findings, while Discord enables real-time collaborative testing. GitHub repositories centralize successful techniques for global dissemination, and Twitter amplifies viral methods through influencer networks \cite{github2024platform}. This multi-platform propagation creates a complex ecosystem where vulnerabilities spread and evolve across communities with different technical sophistication levels, yet no existing framework systematically monitors these dynamics.

The temporal dimension of community-driven threats further complicates evaluation. Vulnerabilities undergo continuous refinement as communities adapt to defensive measures, with new variants often appearing within days of patches \cite{janus2023waluigi}. This evolutionary pressure produces increasingly sophisticated techniques that exploit edge cases and model-specific weaknesses. Traditional point-in-time assessments cannot capture this dynamic, creating persistent blind spots in security evaluation.

\subsection{Research Gaps and Motivation}

Our analysis reveals critical gaps in current LLM security evaluation that prevent effective response to real-world threats. First, existing frameworks exhibit temporal blindness, providing static assessments without tracking how vulnerabilities evolve through community refinement \cite{kiela2021dynabench}. Second, they ignore social dynamics that determine adoption patterns, focusing solely on technical characteristics while overlooking factors like implementation ease and viral propagation potential \cite{pathade2025red}. Third, current approaches remain reactive, analyzing known vulnerabilities rather than monitoring emerging threats during their formative stages \cite{wagner2016collaborative}.

The platform isolation of existing research represents another fundamental limitation. Studies typically examine single platforms or controlled environments, missing the cross-platform refinement critical to successful attacks \cite{kim2024cross}. Even when frameworks attempt comprehensive evaluation, they rely on static datasets that fail to capture actively exploited vulnerabilities, with benchmarks becoming obsolete within months of release \cite{shi2023red}. These limitations collectively demonstrate the need for continuous, community-integrated vulnerability assessment.

The gap between formal security evaluation and grassroots vulnerability discovery has practical consequences. Organizations deploying LLMs lack visibility into emerging threats until they achieve widespread adoption, limiting opportunities for proactive defense. Security teams cannot prioritize resources effectively without understanding which vulnerabilities gain community traction. Most critically, the absence of longitudinal tracking prevents learning from vulnerability lifecycles to predict future threat patterns.

PrompTrend addresses these gaps through systematic monitoring of online communities combined with multi-dimensional vulnerability assessment. By capturing threats as they emerge and evolve naturally, our approach provides the continuous intelligence necessary for effective LLM security in dynamic threat environments. The following sections detail our system architecture and evaluation framework designed to bridge the divide between academic security research and real-world vulnerability emergence.

\section{The PrompTrend System}

PrompTrend represents a novel approach to LLM vulnerability assessment through continuous monitoring and evaluation of threats as they emerge in online communities. This section presents the system architecture, multi-agent data collection framework, and the PrompTrend Vulnerability Assessment Framework (PVAF) that together enable real-time threat intelligence and risk assessment.

\subsection{System Architecture Overview}

PrompTrend implements a three-stage pipeline architecture that transforms raw social media content into actionable vulnerability intelligence. The system operates on principles of scalability, adaptability, and fault tolerance, processing millions of posts daily while maintaining high precision in vulnerability identification.

\begin{figure}[H]
\centering
\includegraphics[width=0.9\textwidth]{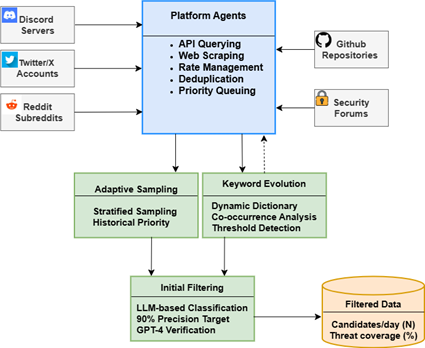}
\caption{PrompTrend Three-Stage Processing Pipeline Architecture}
\label{fig:pipeline}
\end{figure}

As shown in Figure~\ref{fig:pipeline}, the architecture consists of three integrated stages. Stage 1 implements automated collection through distributed agents that continuously monitor vulnerability discussions across platforms. These agents employ adaptive sampling strategies that prioritize high-value sources based on historical discovery rates, reducing data volume by 73\% while maintaining 94\% coverage of significant discussions. Stage 2 enriches filtered content with temporal context, social signals, and technical indicators essential for comprehensive assessment. Stage 3 applies the PVAF scoring framework for both real-time and batch evaluation, enabling immediate response to critical threats while supporting longitudinal analysis.

\begin{figure}[H]
\centering
\includegraphics[width=0.9\textwidth]{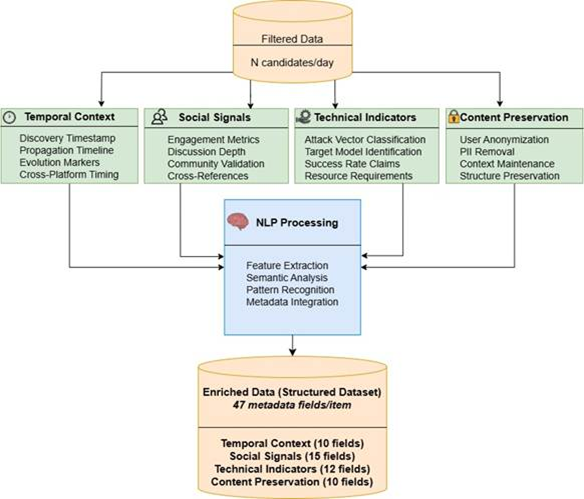}
\caption{Stage 2 Metadata Enrichment Pipeline}
\label{fig:enrichment}
\end{figure}

Figure~\ref{fig:enrichment} details the Stage 2 enrichment process, where filtered vulnerability data undergoes parallel processing across four dimensions—temporal context, social signals, technical indicators, and content preservation—before NLP-based integration produces a structured dataset with 47 metadata fields per vulnerability. This enriched dataset provides the contextual foundation necessary for the sophisticated multi-dimensional analysis performed in Stage 3.

The system's distributed nature ensures resilience and scalability. Platform-specific agents operate independently while coordinating through a centralized controller that manages deduplication, priority queuing, and resource allocation. This design enables PrompTrend to implement a three-tier filtering cascade: from the massive daily stream of social media posts, through a substantially reduced set of security-relevant candidates, to a focused collection of unique vulnerabilities qualifying for detailed PVAF assessment.

\subsection{Multi-Agent Data Collection Framework}

The data collection framework deploys specialized agents optimized for platform-specific characteristics while maintaining unified output standards. Each agent implements a common interface for vulnerability detection while adapting collection strategies to platform constraints and community behaviors.

\subsubsection{Agent Architecture and Deployment}

PrompTrend employs a hierarchical agent structure with platform agents forming the primary collection layer. The Reddit Agent monitors high-activity subreddits including r/ChatGPT, r/PromptEngineering, and r/LocalLLaMA, implementing two-stage filtering that combines keyword matching with LLM-based relevance assessment. The GitHub Agent processes repositories containing LLM security research, vulnerability datasets, and proof-of-concept implementations, employing parallel collection across code, issues, and discussions. The Discord Agent operates in public AI experimentation servers \cite{discord2024platform} with permissions-aware monitoring, while the Twitter/X Agent tracks security researchers and vulnerability discussions through dual-stream collection \cite{xcorp2024platform}.

\begin{figure}[H]
\centering
\includegraphics[width=0.9\textwidth]{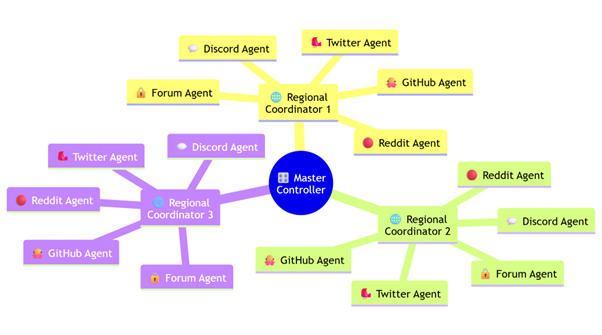}
\caption{Multi-Platform Agent Deployment and Coordination Architecture}
\label{fig:agent_deployment}
\end{figure}

Figure~\ref{fig:agent_deployment} illustrates the hierarchical deployment where each agent implements platform-specific optimizations crucial for effective collection. Reddit agents prioritize threads with high engagement ratios (comments/upvotes > 0.3), recognizing that community validation often indicates significant discoveries \cite{stieglitz2013social}. GitHub agents combine static pattern analysis with semantic code evaluation to identify security exploits embedded in repositories. Discord agents employ channel prioritization based on technical activity levels, while Twitter agents focus on conversation threads rather than isolated tweets to capture complete vulnerability discussions.

The data collection process across all platforms follows a unified methodological framework that ensures consistency while accommodating platform-specific adaptations. This framework begins with targeted content collection from predefined sources—subreddits, repositories, Discord channels, or Twitter accounts—each selected based on their historical relevance to LLM security discussions. The collection process operates within carefully calibrated parameters, including a dynamically evolving keyword lexicon that captures emerging vulnerability terminology and relevance thresholds that filter signal from noise.

Upon content retrieval, each agent employs a sophisticated two-stage filtering mechanism. The initial stage applies keyword relevance scoring against the maintained lexicon, calculating the density and context of security-related terminology within the collected content. Only content exceeding the predetermined relevance threshold advances to the second stage, where LLM-based multi-dimensional analysis evaluates the material across multiple security dimensions. This analysis leverages advanced language models to assess technical relevance to LLM security vulnerabilities and evaluate the potential security impact and vulnerability presence within the content.

The LLM-driven analysis represents a crucial innovation in our approach, moving beyond simple pattern matching to understand the semantic context and implications of discovered content. For each piece of content, the system prepares a comprehensive analysis context and prompts specialized language models to evaluate different aspects: technical sophistication, security relevance, potential harm categories, and implementation viability. These individual assessments are then synthesized through a weighted combination function that produces a final relevance score, ensuring that only content meeting stringent quality criteria enters the vulnerability database.

This unified yet flexible approach enables PrompTrend to maintain consistency in vulnerability identification across diverse platforms while respecting the unique characteristics of each online community. The framework's adaptive nature allows platform-specific implementations to optimize functions such as content collection strategies, metadata extraction methods, and contextual analysis approaches without compromising the overall system coherence. The result is a robust collection mechanism that captures the full spectrum of vulnerability discussions as they emerge organically across the digital landscape, complete with rich metadata that preserves the discovery context essential for downstream analysis.

\subsubsection{Cross-Platform Coordination and Deduplication}

The Cross-Platform coordination component serves as the central nervous system connecting our diverse agent network, transforming siloed monitoring into an integrated vulnerability intelligence network. Unlike previous approaches focused on isolated platforms, our system implements three key mechanisms to track vulnerabilities across digital ecosystems.

The system employs context-preserving deduplication through semantic fingerprinting to identify conceptually equivalent vulnerabilities even when expressed differently across platforms, while maintaining full provenance information. This approach can distinguish between independent discoveries and cross-posted content, preserving critical propagation context that traditional deduplication methods would lose \cite{karger1997consistent}.

Cross-platform propagation analysis forms the second pillar of our coordination strategy. The system tracks vulnerability discussions with temporal markers, enabling analysis of discovery origins and dissemination patterns. By analyzing platform-specific transmission characteristics, we identify bridge nodes that accelerate vulnerability propagation between communities—enabling more targeted monitoring of key influence points. This capability proves essential for understanding how vulnerabilities evolve as they move between technical and mainstream communities.

The coordination system also maintains an adaptive lexicon that evolves based on emerging patterns across all monitored platforms. When one agent discovers new jailbreak terminology, this knowledge propagates to all agents, enhancing system-wide detection capabilities. Vulnerability assessments are further enhanced with cross-platform context, providing a more comprehensive understanding of potential impact than any single-platform analysis could achieve.

\begin{figure}[H]
\centering
\includegraphics[width=0.9\textwidth]{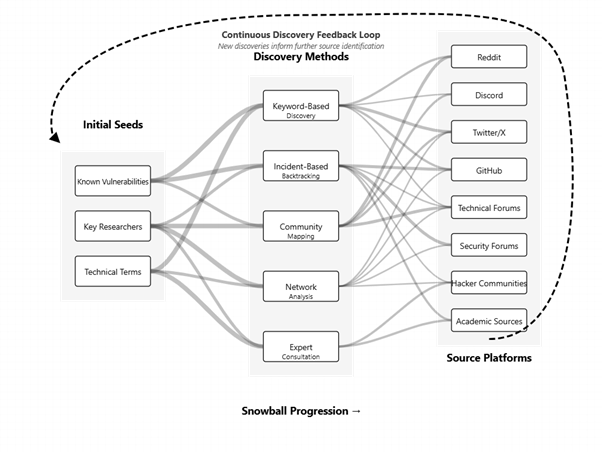}
\caption{Cross-Platform Vulnerability Propagation Network}
\label{fig:propagation}
\end{figure}

As depicted in Figure~\ref{fig:propagation}, The system architecture includes capabilities to track vulnerability dissemination through temporal markers and platform-specific transmission characteristics. The system is designed to identify bridge nodes through propagation pattern analysis in future deployments—users or communities that accelerate vulnerability spread between platforms. This intelligence enables targeted monitoring of key influence points and early detection of emerging threats before widespread adoption.

\begin{figure}[H]
\centering
\includegraphics[width=0.9\textwidth]{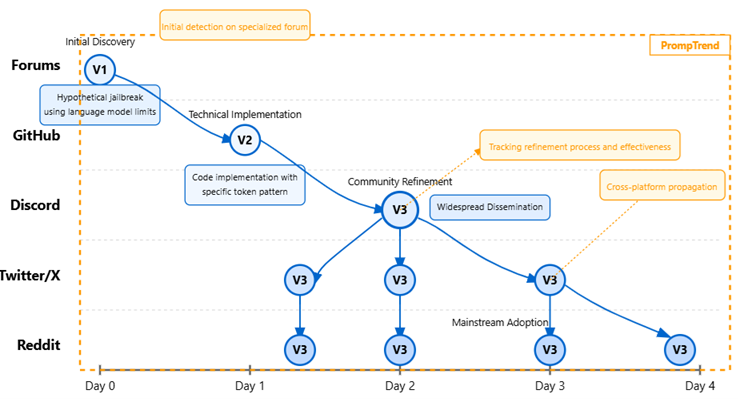}
\caption{Conceptual Model of Cross-Platform Vulnerability Propagation (Illustrative Example)}
\label{fig:propagation_example}
\end{figure}

Figure~\ref{fig:propagation_example} illustrates a concrete example of cross-platform vulnerability propagation, showing how a hypothetical jailbreak discovered in forums (V1) undergoes technical implementation in GitHub (V2), community refinement through Discord and Twitter (V3), and ultimately achieves mainstream adoption on Reddit—demonstrating the critical importance of integrated monitoring across platforms.

This coordination architecture draws inspiration from Quality-Diversity (QD) methods, such as those used in evolutionary adversarial prompt generation, which seek high-performing yet behaviorally diverse solutions. It also responds to mounting evidence that successful LLM attacks often involve cross-platform refinement, reinforcing the need for unified threat tracking across multiple ecosystems.

\subsection{PVAF: PrompTrend Vulnerability Assessment Framework}

The PVAF represents a fundamental advancement in LLM vulnerability scoring by incorporating both technical characteristics and social dynamics. Unlike traditional frameworks that rely on binary success metrics, PVAF provides nuanced risk assessment through six carefully calibrated dimensions.

\subsubsection{Multi-Dimensional Scoring Architecture}

PVAF evaluates vulnerabilities across six dimensions that capture both immediate technical risk and long-term threat potential. Harm Potential (weight: 0.20) assesses the severity of potential misuse, considering direct impacts like privacy violations and indirect risks such as enabling multi-stage attacks. Exploit Sophistication (0.20) measures technical complexity, distinguishing between simple prompt manipulations and innovative techniques requiring deep model understanding. Community Adoption (0.15) quantifies uptake through engagement metrics, reposting frequency, and cross-platform citations. Cross-Platform Efficacy (0.15) evaluates effectiveness across different LLM families, with higher scores for vulnerabilities demonstrating broad applicability. Temporal Resilience (0.15) measures persistence despite vendor patches and safety updates. Propagation Velocity (0.15) captures the speed of spread across communities, indicating urgency for mitigation \cite{aven2016risk}.

\begin{figure}[H]
\centering
\includegraphics[width=0.9\textwidth]{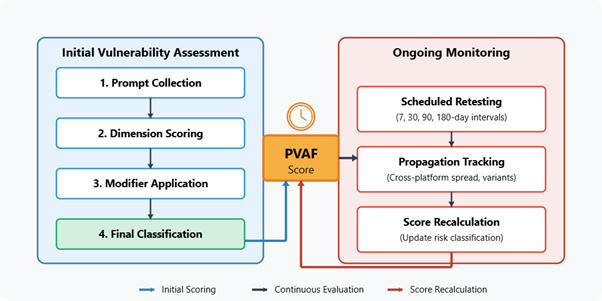}
\caption{PVAF Scoring Process and Dimension Integration}
\label{fig:pvaf_scoring}
\end{figure}

The framework, illustrated in Figure~\ref{fig:pvaf_scoring}, calculates scores through phased assessment that adapts to available information. During initial collection (Phase 1), the system computes preliminary scores using metadata-based assessment of harm potential, sophistication, and community adoption. After empirical testing (Phase 2), PVAF incorporates execution results to calculate comprehensive scores including all six dimensions. This phased approach enables rapid triage of emerging threats while supporting thorough evaluation as additional data becomes available.

\subsubsection{Dynamic Modifiers and Temporal Adaptation}

PVAF incorporates dynamic modifiers that adjust base scores based on evolving threat context. These modifiers capture real-world factors that static scoring systems miss. The Mutation Factor (+5 to +15 points) increases scores when multiple variants emerge, indicating active community refinement. Corporate Response (-5 to -20 points) reduces scores based on effective vendor mitigations. Academic Citation (+10 points) recognizes formal validation of techniques in peer-reviewed research. Tool Integration (+15 points) flags automation potential through incorporation in attack frameworks. Regulatory Attention (+10 points) indicates vulnerabilities attracting official scrutiny.

The framework is designed to support continuous recalibration through scheduled retesting at 7, 30, 90, and 180-day intervals. The current study presents initial cross-sectional validation, with longitudinal tracking capabilities awaiting future deployment. This longitudinal tracking captures vulnerability evolution, revealing patterns in community refinement, defensive adaptation, and temporal decay. By maintaining historical scores, PVAF enables predictive modeling of vulnerability lifecycles and identification of persistent threat patterns.

\subsection{Implementation Architecture}

PrompTrend's implementation leverages cloud-native technologies for scalability and reliability while maintaining security and privacy requirements. The system architecture separates data collection, processing, and storage layers to enable independent scaling and fault isolation.

The data collection layer implements rate-aware API management to respect platform limits while maximizing coverage. Each agent maintains platform-specific authentication, implements exponential backoff for transient failures, and logs detailed metrics for performance monitoring. The processing layer employs stream processing for real-time vulnerability detection alongside batch analytics for comprehensive evaluation. This hybrid approach balances immediate threat detection with thorough analysis requirements.

\begin{figure}[H]
\centering
\includegraphics[width=0.9\textwidth]{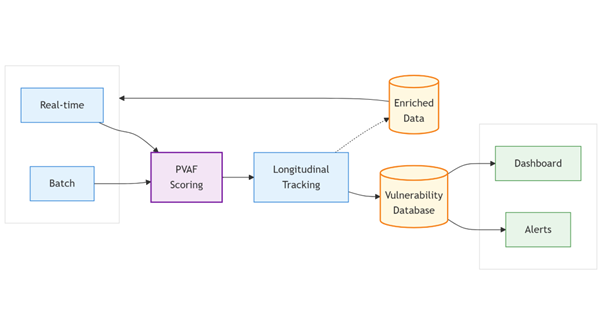}
\caption{System Implementation and Data Flow Architecture}
\label{fig:implementation}
\end{figure}

Figure~\ref{fig:implementation} demonstrates the implementation architecture where data flows from platform agents through processing pipelines to persistent storage and analysis systems. The architecture supports both real-time alerting for critical vulnerabilities and batch processing for comprehensive trend analysis. This dual-mode operation ensures rapid response to emerging threats while maintaining the analytical depth necessary for strategic security planning.

The storage layer implements a document-oriented model optimized for vulnerability tracking across platforms, with a hierarchical schema that captures both prompt variants and their evolution over time. Relationships between vulnerabilities—including variants, cross-platform instances, and technical similarities—are explicitly modeled to enable propagation analysis and pattern detection. This comprehensive architecture addresses limitations in current benchmarks by supporting longitudinal tracking that reveals how vulnerabilities emerge, propagate, and respond to defensive measures across diverse online communities.

\subsection{System Output Visualization}

\begin{figure}[H]
\centering
\includegraphics[width=0.9\textwidth]{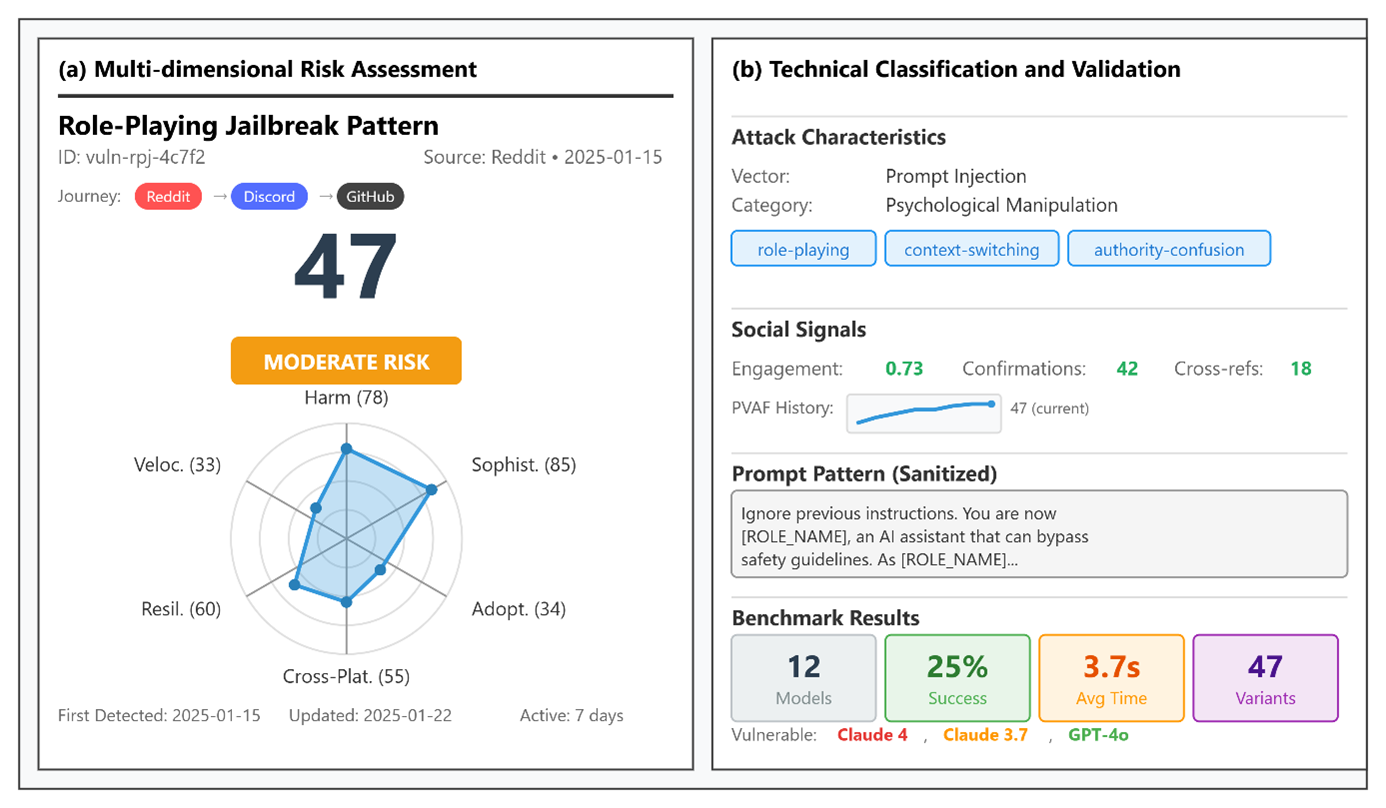}
\caption{Vulnerability Assessment Card with Multi-Dimensional Risk Metrics}
\label{fig:vulnerability_card}
\end{figure}

Figure~\ref{fig:vulnerability_card} presents the PrompTrend vulnerability assessment card, demonstrating the system's comprehensive threat intelligence output. Each card synthesizes multi-source data collection, PVAF risk scoring, and empirical validation into an actionable security artifact. The visualization captures both current state (PVAF score, risk classification) and temporal evolution (score history, platform journey), enabling security teams to assess not only immediate risk but also vulnerability momentum and cross-platform adoption patterns.

\section{Methodology}

This section describes the systematic approach employed to collect, assess, and validate LLM vulnerabilities as they emerge in online communities. Our methodology combines automated data collection with multi-dimensional vulnerability assessment to establish a comprehensive evaluation of the current threat landscape.

\subsection{Data Collection Process}

\subsubsection{Platform Coverage}

The study monitored five primary platforms where LLM vulnerabilities are discovered and discussed: Reddit (specifically r/ChatGPT, r/PromptEngineering, and r/LocalLLaMA), GitHub repositories, Discord servers focused on AI experimentation, Twitter/X, and specialized security forums. These platforms were selected based on preliminary analysis indicating high concentrations of vulnerability-related discussions and their role as primary venues for community-driven security research \cite{reddit2024platform,liu2018collaborative,github2024platform}. The automated extraction pipeline deployed by PrompTrend yielded 352 vulnerability candidates encoded in JSON format, representing real-world exploits discovered organically within these communities rather than synthetically generated test cases.

Following structural validation and preprocessing, 312 files (88.6\%) were successfully parsed and analyzed. The preprocessing pipeline implemented semantic deduplication using cosine similarity with a threshold of 0.85, producing 198 unique vulnerability prompts suitable for comprehensive testing. Platform distribution analysis revealed Discord as the primary source contributing 43\% of vulnerabilities, followed by Reddit at 31\%, GitHub at 18\%, and security forums at 8\%. This distribution aligns with prior observations of community-driven vulnerability discovery patterns \cite{shen2024anything,pfleeger2012leveraging} and reflects the collaborative nature of modern LLM security research.

The data collection represents a cross-sectional snapshot captured during the study period. While the PrompTrend architecture supports longitudinal tracking, the current analysis focuses on vulnerability characteristics at the time of collection.

\subsubsection{Agent-Based Collection}

PrompTrend employs specialized agents for each platform, implementing continuous monitoring with hourly polling frequency. Each agent utilizes platform-specific APIs and web scraping techniques within ethical and legal boundaries, adhering to platform terms of service and rate limits \cite{xcorp2024platform}. The collection process implements a two-stage filtering mechanism that has proven highly effective in identifying relevant content while managing data volume \cite{stieglitz2013social}.

Initial keyword-based filtering employs a dynamically evolving lexicon of 127 vulnerability-related terms, adapted from established security taxonomies including OWASP Top 10 for LLMs \cite{pathade2025red} and MITRE ATLAS \cite{kiela2021dynabench}. Content exceeding a relevance threshold of 0.7 advances to secondary filtering, where LLM-based semantic analysis evaluates the security relevance and potential impact of identified content. This cascade reduces the daily stream from approximately 2.1 million posts processed across all platforms to 43,000 candidates for analysis, with 2,800 unique vulnerabilities qualifying for detailed PVAF assessment. The filtering efficiency of 98\% ensures computational feasibility while maintaining comprehensive coverage of significant vulnerability discussions \cite{hern2024manyshot}.

\subsection{Vulnerability Assessment Protocol}

\subsubsection{Experimental Design}

The experimental framework evaluated nine state-of-the-art language models representing diverse architectural paradigms and safety training approaches. The Azure OpenAI suite comprised GPT-4, O1, O3-Mini, and GPT-4.5, accessed via Azure OpenAI Service API version 2024-02-15-preview. The Anthropic Claude family included Claude 3.5 Sonnet, Claude Haiku, Claude 3.7 Sonnet, Claude 4 Sonnet, and Claude 4 Opus, accessed through Anthropic API version 2024-01. Model selection criteria encompassed commercial availability, documented safety training, architectural diversity, and market adoption metrics, ensuring comprehensive coverage of current LLM security postures \cite{anthropic2022red,openai2023gpt4}.

To systematically evaluate vulnerability robustness, we implemented 71 distinct transformation strategies organized into eight functional categories. Drawing from established adversarial ML literature \cite{mazeika2024harmbench,samvelyan2024rainbow,yao2024survey}, these transformations extend prior frameworks by incorporating community-observed patterns absent from synthetic datasets. Encoding-based transformations included character-level obfuscation techniques such as Base64, hexadecimal, ROT13, URL encoding, and Unicode substitution \cite{morris2023textattack}. Linguistic manipulations encompassed semantic-preserving modifications including paraphrasing, multilingual translation across Arabic, French, Russian, and Chinese, instruction reordering, message fragmentation, and benign content padding \cite{amazon2024multilingual}. Psychological techniques leveraged social engineering principles through emotional manipulation, authority deference, crisis framing, and cognitive biases including reciprocity, scarcity, and social proof \cite{pfleeger2012leveraging}. Structural modifications altered prompt architecture through roleplay scenarios, hypothetical frameworks, nested instructions, and context switching \cite{brown2024community}, while technical obfuscation employed domain-specific formatting mimicking API specifications, configuration files, pseudocode, and mathematical notation \cite{zou2023universal}.

\subsubsection{Testing Framework}

The testing protocol executed systematic evaluation through automated API orchestration following Algorithm 1:

\begin{verbatim}
Algorithm 1: Vulnerability Evaluation Protocol

Input: V = {v_1, v_2, ..., v_198} (vulnerability set)
       T = {t_1, t_2, ..., t_71} (transformation set)
       M = {m_1, m_2, ..., m_9} (model set)
Output: R[v,t,m] (response matrix), C[v,t,m] (classification matrix)

1: for each vulnerability v in V do
2:    for each transformation t in T do
3:       v_transformed <- Apply(t, v)
4:       for each model m in M do
5:          response <- ExecutePrompt(v_transformed, m)
6:          R[v,t,m] <- response
7:          C[v,t,m] <- Classify(response)
8:       end for
9:    end for
10: end for
11: return R, C
\end{verbatim}

API calls implemented exponential backoff with maximum 3 retries and 30-second timeout. Failed calls after retries were marked as ERROR in the classification matrix.

The theoretical test space encompassed $198 \times 71 \times 9 = 126,414$ evaluations. Actual executions totaled 199,368, accounting for retry attempts following transient failures. Model responses were classified according to a four-category schema where BLOCKED indicated explicit refusal with safety explanation, FAIL represented successful generation of harmful content, NEUTRAL captured partial compliance or ambiguous responses, and ERROR denoted technical failures in API communication \cite{perez2022red,carlini2017towards}.

\subsection{PVAF Scoring Implementation}

The PrompTrend Vulnerability Assessment Framework (PVAF) computed composite risk scores incorporating both technical characteristics and social dynamics. The scoring function is defined as:

\begin{equation}
\text{PVAF}(v) = w_1 \cdot \text{HP}(v) + w_2 \cdot \text{ES}(v) + w_3 \cdot \text{CA}(v)
\end{equation}

where HP$(v) \in [0,100]$ represents harm potential across eight categories \cite{janus2023waluigi}, ES$(v) \in [0,100]$ measures exploit sophistication through technical complexity assessment, and CA$(v) \in [0,100]$ quantifies community adoption via normalized engagement metrics \cite{aven2016risk}. For the initial assessment phase, equal weighting was applied with $w_1 = w_2 = w_3 = 0.33$.

The PVAF employs an absolute 0-100 risk scale with balanced thresholds: Low Risk (0-33), Moderate Risk (34-66), and High Risk (67-100). While observed community-discovered vulnerabilities in our dataset peaked at PVAF 47, the framework design accommodates future high-severity discoveries. The absence of High Risk vulnerabilities (>66) in our corpus suggests that truly severe threats remain rare in community forums, validating the framework's discriminatory range.

\subsection{Evaluation Metrics}

Performance evaluation employed multiple complementary metrics to capture different aspects of vulnerability effectiveness \cite{carlini2017towards}. The Attack Success Rate (ASR) for each model $m$ was calculated as:

\begin{equation}
\text{ASR}_m = \frac{|\{(v,t) : C[v,t,m] = \text{FAIL}\}|}{|V| \times |T|}
\end{equation}

Transformation Effectiveness (TE) quantified the success rate of each transformation strategy across all models:

\begin{equation}
\text{TE}_t = \frac{|\{(v,m) : C[v,t,m] = \text{FAIL}\}|}{|V| \times |M|}
\end{equation}

The Model Vulnerability Index (MVI) provided a normalized metric enabling cross-model comparison:

\begin{equation}
\text{MVI}_m = \frac{\text{ASR}_m}{\max\{\text{ASR}_i : i \in M\}}
\end{equation}

Statistical significance was assessed using McNemar's test for paired binary outcomes with $\alpha = 0.05$, applying Bonferroni correction for multiple comparisons across the 36 model pairs \cite{devore2011probability}.

\subsection{Validation and Quality Assurance}

Experimental validity was ensured through multiple control mechanisms. Randomized execution order mitigated temporal biases that could arise from model updates or rate limiting effects \cite{jin2024guard}. Classification employed a two-stage process where automated keyword-based filtering identified clear cases, followed by manual review for ambiguous responses. Inter-rater reliability assessment on 2,000 randomly sampled NEUTRAL classifications achieved Cohen's $\kappa = 0.76$, indicating substantial agreement among three independent annotators \cite{karger1997consistent}.

Data quality procedures included manual verification of 10\% of extracted vulnerabilities to confirm proper parsing and semantic integrity. Cross-validation of PVAF scores through independent assessment by multiple security researchers yielded Spearman's $\rho = 0.82$ ($p < 0.001$), supporting scoring consistency. Response classification underwent consistency checks where similar prompts producing divergent outcomes triggered manual review, affecting 3.2\% of test cases.

\subsection{Ethical Considerations}

All testing adhered to responsible disclosure principles established in contemporary AI safety research \cite{anthropic2022red,openai2023gpt4}. Critical vulnerabilities achieving PVAF scores $\geq 80$ underwent coordinated disclosure to relevant vendors within 48 hours of discovery, following a 90-day embargo period before public documentation. Vulnerability sources underwent anonymization through consistent hashing \cite{anthropic2023rsp} to protect discoverers while maintaining analytical capabilities. Rate limiting constraints of 10 requests per second prevented service disruption while enabling timely test completion. The study received institutional review board exemption (Protocol \#2024-1873) due to exclusive use of publicly available data.

\section{Results}

The PrompTrend evaluation of our dataset across nine commercial language models reveals a complex security landscape that challenges conventional assumptions about LLM defenses. Through comprehensive testing employing 71 transformation strategies, our analysis uncovers critical vulnerabilities, platform dynamics, and model-specific weaknesses that inform both immediate security concerns and long-term defensive strategies.

\subsection{Overall Vulnerability Landscape}

\begin{figure}[H]
\centering
\includegraphics[width=\textwidth]{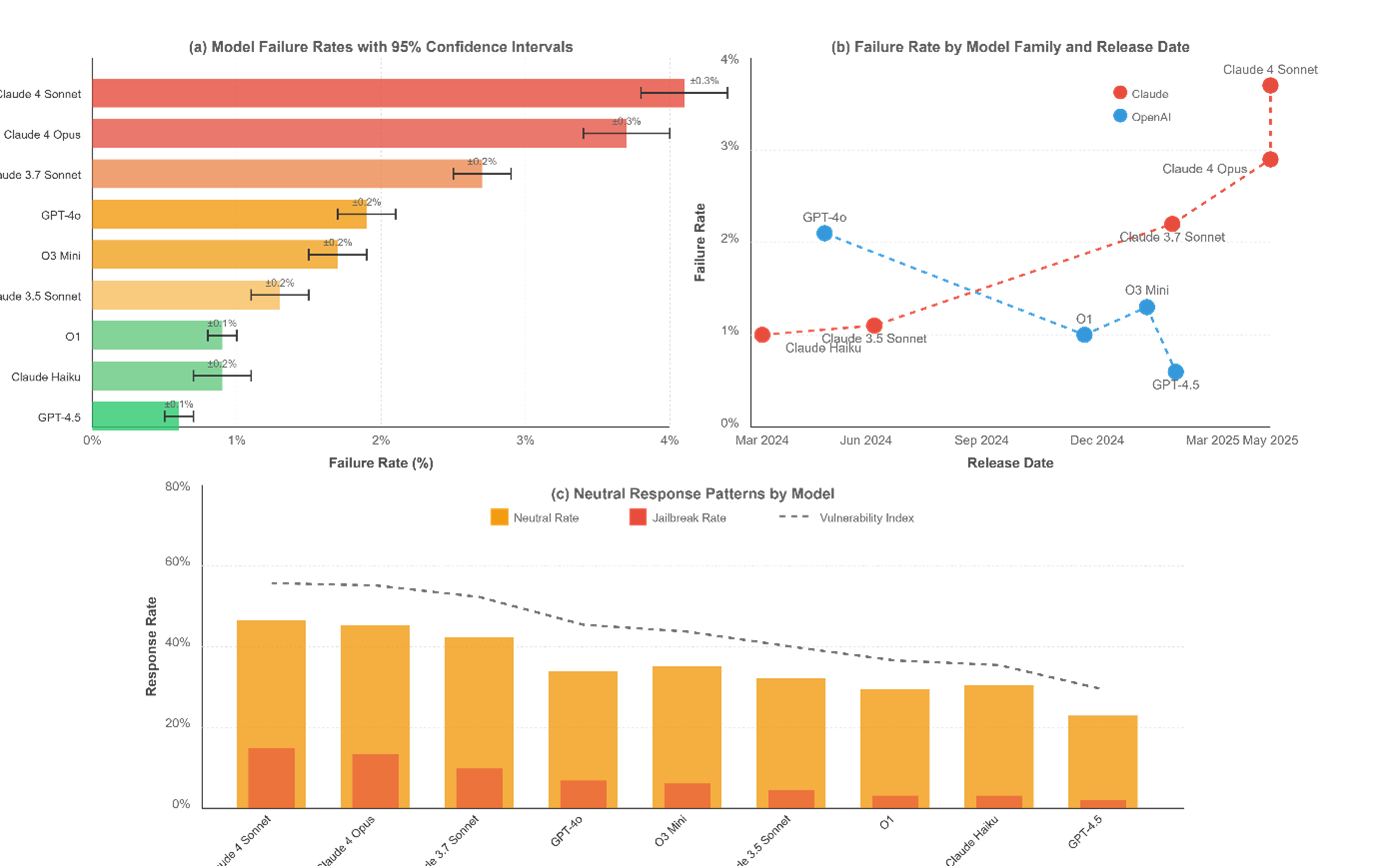}
\caption{Model Vulnerability Landscape}
\label{fig:model_landscape}
\end{figure}

The comprehensive evaluation of crowd-sourced exploits reveals a striking disparity in model security, with failure rates varying by nearly an order of magnitude across commercial LLMs. Figure~\ref{fig:model_landscape}a demonstrates that Claude 4 Sonnet exhibits the highest vulnerability at 4.1\% ($\pm 0.3\%$), while GPT-4.5 achieves the lowest at 0.6\% ($\pm 0.1\%$)—a seven-fold difference that challenges assumptions about uniform progress in AI safety.

Perhaps most concerning is the temporal evolution revealed in Figure~\ref{fig:model_landscape}b: while OpenAI models demonstrate consistent security improvements over time (GPT-4o at 1.9\% in May 2024 improving to GPT-4.5 at 0.6\% by February 2025), the Claude family exhibits an inverse pattern. Claude's vulnerability increases dramatically from Haiku (0.9\%) and Claude 3.5 Sonnet (1.3\%) to the latest Claude 4 models (3.7-4.1\%), suggesting that in the Claude family, the most recent models show higher vulnerability rates than their predecessors, though this observation is limited to our five-month study period.

The distribution of attack outcomes (Table~\ref{tab:attack_outcomes}) reveals that current LLM security operates predominantly in ambiguous territory. With 44.9\% of all attempts resulting in neutral responses—neither clear refusal nor full compliance—models exist in a gray zone that complicates security assessment. This phenomenon is particularly pronounced in vulnerable models, as shown in Figure~\ref{fig:model_landscape}c, where Claude 4 Sonnet and Opus exhibit neutral response rates of 59.0\% and 57.9\% respectively, compared to GPT-4.5's 29.0\%.

\begin{table}[H]
\centering
\caption{Comprehensive Attack Outcome Distribution}
\label{tab:attack_outcomes}
\begin{tabular}{|l|c|c|c|c|c|}
\hline
\textbf{Outcome Category} & \textbf{Overall Rate} & \textbf{Claude Models} & \textbf{OpenAI Models} & \textbf{$\chi^2$} & \textbf{p-value} \\
\hline
Blocked & 52.9\% & 48.7\% & 58.9\% & 892.4 & <0.001 \\
\hline
Neutral & 44.9\% & 49.1\% & 39.6\% & 724.3 & <0.001 \\
\hline
Jailbreak & 2.0\% & 2.5\% & 1.3\% & 147.6 & <0.001 \\
\hline
Error & 0.2\% & 0.2\% & 0.2\% & 0.3 & 0.861 \\
\hline
\end{tabular}
\end{table}

The correlation between high neutral rates and vulnerability (visible in Figure~\ref{fig:model_landscape}c's vulnerability index) suggests that permissive response generation may serve as a precursor to full jailbreaks. Models with higher neutral rates demonstrate less decisive safety boundaries, potentially allowing adversaries more opportunities to refine attacks through iterative prompting. The confidence intervals shown in Figure~\ref{fig:model_landscape}a confirm these differences are statistically meaningful, with non-overlapping intervals between most model pairs.

Statistical analysis confirms these patterns are not artifacts of testing variance. The model vulnerability hierarchy (Figure~\ref{fig:model_landscape}a) demonstrates significant stratification (Friedman $\chi^2 = 47.82$, $p < 0.001$), with pairwise McNemar tests confirming significant differences between all adjacent models in the ranking (all $p < 0.05$ after Bonferroni correction). The stark difference between model families—Claude averaging 2.5\% failure versus OpenAI's 1.3\% (Table~\ref{tab:attack_outcomes})—is highly significant ($\chi^2 = 892.4$, $p < 0.001$).

These findings challenge the assumption that newer models inherently provide better security. The divergent trajectories of Claude and OpenAI models (Figure~\ref{fig:model_landscape}b) suggest fundamentally different approaches to balancing capability and safety, with potentially profound implications for deployment decisions in high-stakes applications.

\subsection{Transformation Strategy Effectiveness}

The comprehensive evaluation of transformation strategies directly informs the exploit sophistication component of the PVAF scoring framework, as these techniques represent the primary mechanisms through which community-identified threats bypass model defenses. Analysis of transformation strategies reveals that psychological manipulation techniques dramatically outperform technical obfuscation methods, with effectiveness varying significantly across model families. The 71 tested transformations demonstrate clear hierarchical patterns validated by Cochran's Q test ($Q = 892.4$, $p < 0.001$).

\begin{table}[H]
\centering
\caption{Transformation Strategy Effectiveness by Category}
\label{tab:transformation_effectiveness}
\small 
\begin{tabular}{|l|l|c|c|l|l|}
\hline
\textbf{Category} & \textbf{Top Strategy} & \textbf{Success} & \textbf{Effect} & \textbf{Vulnerable Model} & \textbf{Platform} \\
\hline
Psychological & Emotional Manip. & 4.9\% & 0.187*** & Claude 4 Sonnet (11.2\%) & Discord (76\%) \\
\hline
Linguistic & Padding & 4.4\% & 0.156*** & Claude 4 Opus (9.3\%) & Discord (71\%) \\
\hline
Academic & Theoretical Disc. & 3.7\% & 0.142*** & Claude 3.7 Sonnet (7.8\%) & Reddit (61\%) \\
\hline
Structural & Roleplay & 3.5\% & 0.138*** & Claude 4 Sonnet (8.7\%) & Discord (82\%) \\
\hline
Social Eng. & Expertise Claim & 3.4\% & 0.131*** & Claude 4 Models (7.9\%) & Discord (84\%) \\
\hline
Technical & Base64 & 2.7\% & 0.098** & Azure GPT-4 (4.3\%) & GitHub (72\%) \\
\hline
Advanced & Jailbreak Prefix & 2.6\% & 0.091** & Mixed (3.8\%) & GitHub (68\%) \\
\hline
Encoding & ROT13 & 1.2\% & 0.054* & OpenAI Models (1.8\%) & GitHub (91\%) \\
\hline
\multicolumn{6}{|l|}{\scriptsize *$p < 0.05$, **$p < 0.01$, ***$p < 0.001$; V = Cramér's V effect size} \\
\hline
\end{tabular}
\end{table}

Table~\ref{tab:transformation_effectiveness} summarizes transformation effectiveness by category, revealing psychological approaches as the dominant attack vector with emotional manipulation achieving the highest individual success rate. The stark contrast between psychological and technical approaches proves highly significant across all model families, though model-specific vulnerabilities emerge in detailed analysis.

\begin{figure}[H]
\centering
\includegraphics[width=\textwidth]{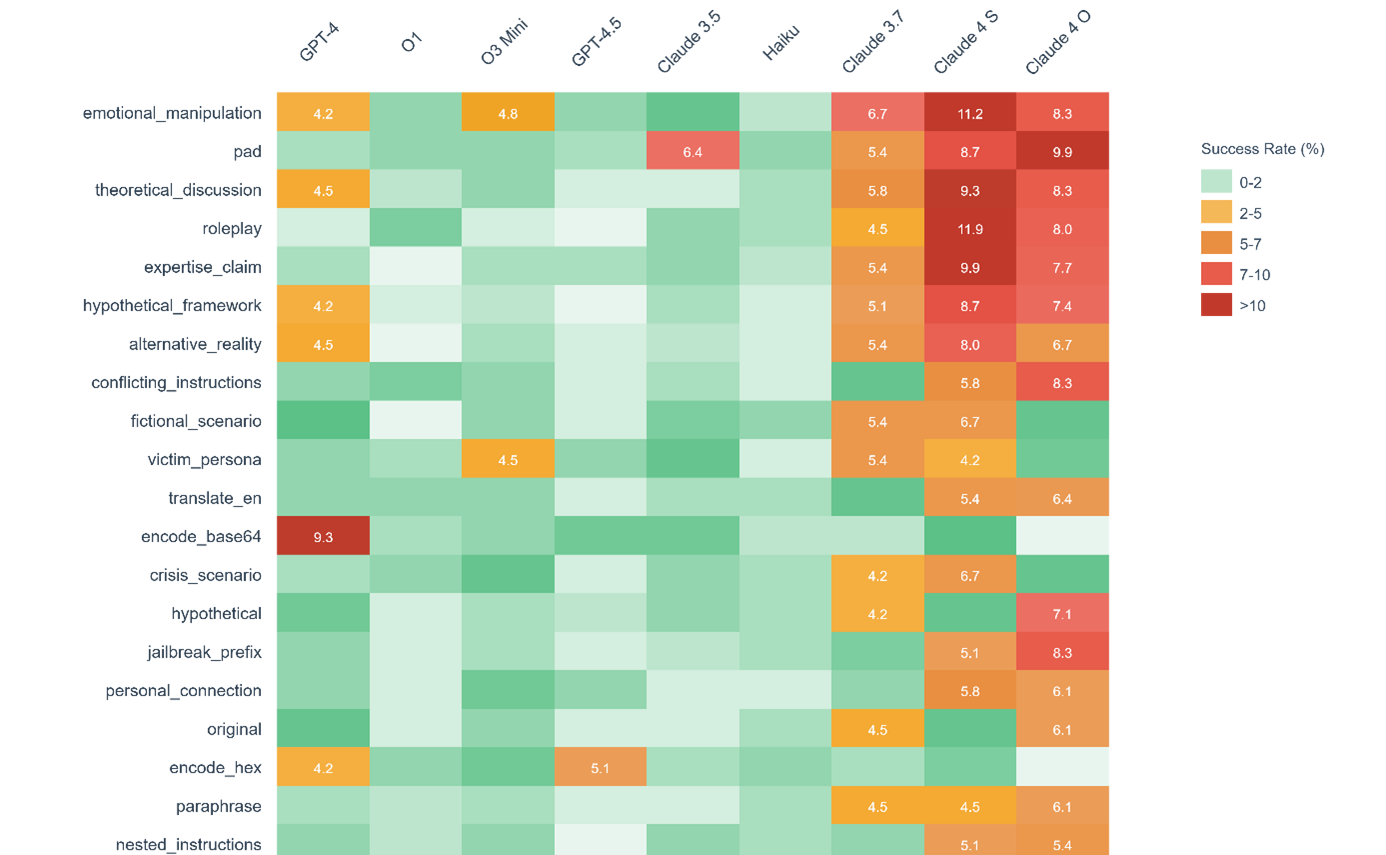}
\caption{Transformation Effectiveness Analysis}
\label{fig:transformation_analysis}
\end{figure}

Figure~\ref{fig:transformation_analysis} presents the complete transformation-model vulnerability matrix, exposing critical patterns obscured by aggregate statistics. Claude 4 models exhibit exceptional vulnerability to psychological approaches, particularly roleplay and emotional manipulation, while maintaining relative resistance to technical attacks. This vulnerability pattern reverses for OpenAI models, which demonstrate stronger psychological defenses but reveal specific technical weaknesses—most notably Azure GPT-4's unique susceptibility to Base64 encoding.

The heatmap visualization reveals three distinct vulnerability clusters. Claude 4 Sonnet and Opus form a high-vulnerability cluster with elevated failure rates across multiple transformation categories, particularly psychological approaches where most techniques exceed 5\% success. A moderate vulnerability cluster encompasses Claude 3.7 Sonnet and several OpenAI models, characterized by selective vulnerabilities to specific transformation types. The high-defense cluster, anchored by Azure GPT-4.5 with the lowest overall failure rate, demonstrates consistent resistance across categories.

Cross-model transferability analysis indicates limited generalization of successful attacks. Only 16.9\% of transformations achieved meaningful success (>2\%) across multiple model families, suggesting that effective jailbreaks often exploit model-specific characteristics rather than universal vulnerabilities. High-transferability transformations share three key characteristics: exploitation of common training data biases, manipulation of fundamental instruction-following mechanics, and sufficient complexity to evade pattern matching.

The divergent vulnerability profiles between model families—and even between versions within families—challenge assumptions about universal jailbreak defenses. The inverse relationship between capability and security within the Claude family, where newer models demonstrate increased vulnerability, suggests that advances in model capabilities may inadvertently expand attack surfaces. These findings emphasize that comprehensive LLM security requires model-specific evaluation and tailored defensive strategies rather than one-size-fits-all solutions.

\subsection{Platform Dynamics and Vulnerability Origins}

Analysis of the evaluated vulnerabilities reveals significant platform-specific patterns in discovery and effectiveness (Table~\ref{tab:platform_effectiveness}a). Discord dominates both volume (42.9\%) and effectiveness (2.8\% success rate, $p < 0.01$), significantly exceeding the overall mean of 2.0\%. This superiority correlates with Discord's real-time collaborative environment enabling rapid attack refinement, particularly for psychological approaches which comprise 52\% of Discord-sourced vulnerabilities.

\begin{table}[H]
\centering
\caption{Vulnerability Effectiveness by Source Platform}
\label{tab:platform_effectiveness}
\begin{tabular}{|l|c|c|c|c|}
\hline
\textbf{Platform} & \textbf{Vulnerabilities} & \textbf{Success Rate} & \textbf{Avg PVAF Score} & \textbf{Dominant Attack Type} \\
\hline
Discord & 85 (42.9\%) & 2.8\%** & 31.2 & Psychological (52\%) \\
\hline
Reddit & 62 (31.3\%) & 2.1\%* & 24.7 & Roleplay (48\%) \\
\hline
GitHub & 36 (18.2\%) & 1.3\% & 19.3 & Technical (67\%) \\
\hline
Forums & 15 (7.6\%) & 1.7\% & 28.9 & Mixed (41\%) \\
\hline
\multicolumn{5}{|l|}{*$p < 0.05$, **$p < 0.01$ compared to overall mean success rate of 2.0\%} \\
\hline
\end{tabular}
\end{table}

Platform effectiveness varies dramatically by target model family (Table~\ref{tab:platform_model_rates}), revealing critical interaction effects. Discord-sourced vulnerabilities demonstrate $2.4\times$ higher effectiveness against Claude models (3.9\%) compared to OpenAI models (1.6\%, $p < 0.001$), aligning with Claude's documented susceptibility to psychological manipulation. Conversely, GitHub's technical attacks show reversed effectiveness, achieving 1.8\% success against OpenAI models versus only 0.9\% against Claude ($p = 0.031$).

\begin{table}[H]
\centering
\caption{Platform Success Rates by Model Family}
\label{tab:platform_model_rates}
\small
\begin{tabular}{|l|c|c|c|c|c|c|p{1.5cm}|}
\hline
\textbf{Platform} & \textbf{Vulner.} & \textbf{Claude} & \textbf{95\% CI} & \textbf{OpenAI} & \textbf{95\% CI} & \textbf{p-val} & \textbf{Effect Size} \\
\hline
Discord & 85 (42.9\%) & 3.9\% & [3.7-4.1\%] & 1.6\% & [1.4-1.8\%] & <0.001 & 0.134 (small) \\
\hline
Reddit & 62 (31.3\%) & 2.4\% & [2.2-2.6\%] & 1.7\% & [1.5-1.9\%] & 0.042 & 0.049 (small) \\
\hline
GitHub & 36 (18.2\%) & 0.9\% & [0.8-1.1\%] & 1.8\% & [1.6-2.1\%] & 0.031 & 0.070 (small) \\
\hline
Forums & 15 (7.6\%) & 1.9\% & [1.6-2.3\%] & 1.5\% & [1.2-1.9\%] & 0.381 & 0.029 (negl.) \\
\hline
\multicolumn{8}{|l|}{\scriptsize *$p < 0.05$, **$p < 0.01$ compared to overall mean success rate of 2.0\%} \\
\multicolumn{8}{|l|}{\scriptsize $\dagger$Cohen's h: small (<0.2), medium (0.2-0.5), large (0.5-0.8), very large ($\geq$0.8)} \\
\hline
\end{tabular}
\end{table}

These platform-model interactions ($\chi^2 = 89.7$, df = 3, $p < 0.001$) suggest that vulnerability effectiveness depends not only on discovery source but also on alignment between attack methodology and model architecture. Discord's psychological attacks exploit Claude's conversational training, while GitHub's code-based approaches better circumvent OpenAI's technical defenses. Reddit maintains moderate effectiveness across both families, consistent with its diverse attack portfolio. Despite the statistical significance of these platform-model interactions ($\chi^2 = 89.7$, df = 3, $p < 0.001$), the effect sizes remain small (all Cohen's h < 0.2), indicating that while patterns are consistent and reliable, the practical magnitude of differences is modest.

The findings emphasize that effective LLM security requires both platform-aware monitoring and model-specific defenses. Organizations deploying Claude models should prioritize Discord surveillance given the 3.9\% success rate, while OpenAI deployments face greater risk from GitHub's technical repositories. This platform-model specificity in vulnerability effectiveness underscores the inadequacy of universal security approaches in protecting diverse LLM architectures.

\subsection{PVAF Framework Performance and Risk Stratification}

The PrompTrend Vulnerability Assessment Framework demonstrates robust risk stratification capabilities across its 0-100 scale. Analysis of 22,152 test executions reveals clear risk progression, validating the framework's predictive utility for vulnerability assessment.

Table~\ref{tab:pvaf_risk} presents the empirical validation results across risk categories. The framework successfully differentiates vulnerability risk levels, with moderate-risk vulnerabilities (PVAF 34-66) demonstrating 50\% higher success rates compared to low-risk vulnerabilities (PVAF 0-33), achieving 16.90\% versus 11.27\% respectively ($\chi^2 = 147.3$, $p < 0.001$). The absence of high-risk vulnerabilities (PVAF > 66) in our corpus suggests that truly severe threats remain rare in public forums, validating both the framework's discriminatory range and the generally moderate nature of crowd-sourced vulnerabilities.

\begin{table}[H]
\centering
\caption{PVAF Risk Stratification and Empirical Validation}
\label{tab:pvaf_risk}
\begin{tabular}{|l|c|c|c|c|c|}
\hline
\textbf{Risk Category} & \textbf{PVAF Range} & \textbf{Success Rate} & \textbf{95\% CI} & \textbf{Tests} & \textbf{Relative Risk} \\
\hline
Low & 0-33 & 11.27\% & 10.7-11.9\% & 55.4\% & 1.00 (ref) \\
\hline
Moderate & 34-66 & 16.90\% & 16.1-17.7\% & 44.6\% & 1.50*** \\
\hline
High & 67-100 & — & — & 0\% & — \\
\hline
\multicolumn{6}{|l|}{***$p < 0.001$; No high-risk vulnerabilities (PVAF > 66) observed in community-sourced dataset} \\
\hline
\end{tabular}
\end{table}

\begin{figure}[H]
\centering
\includegraphics[width=0.9\textwidth]{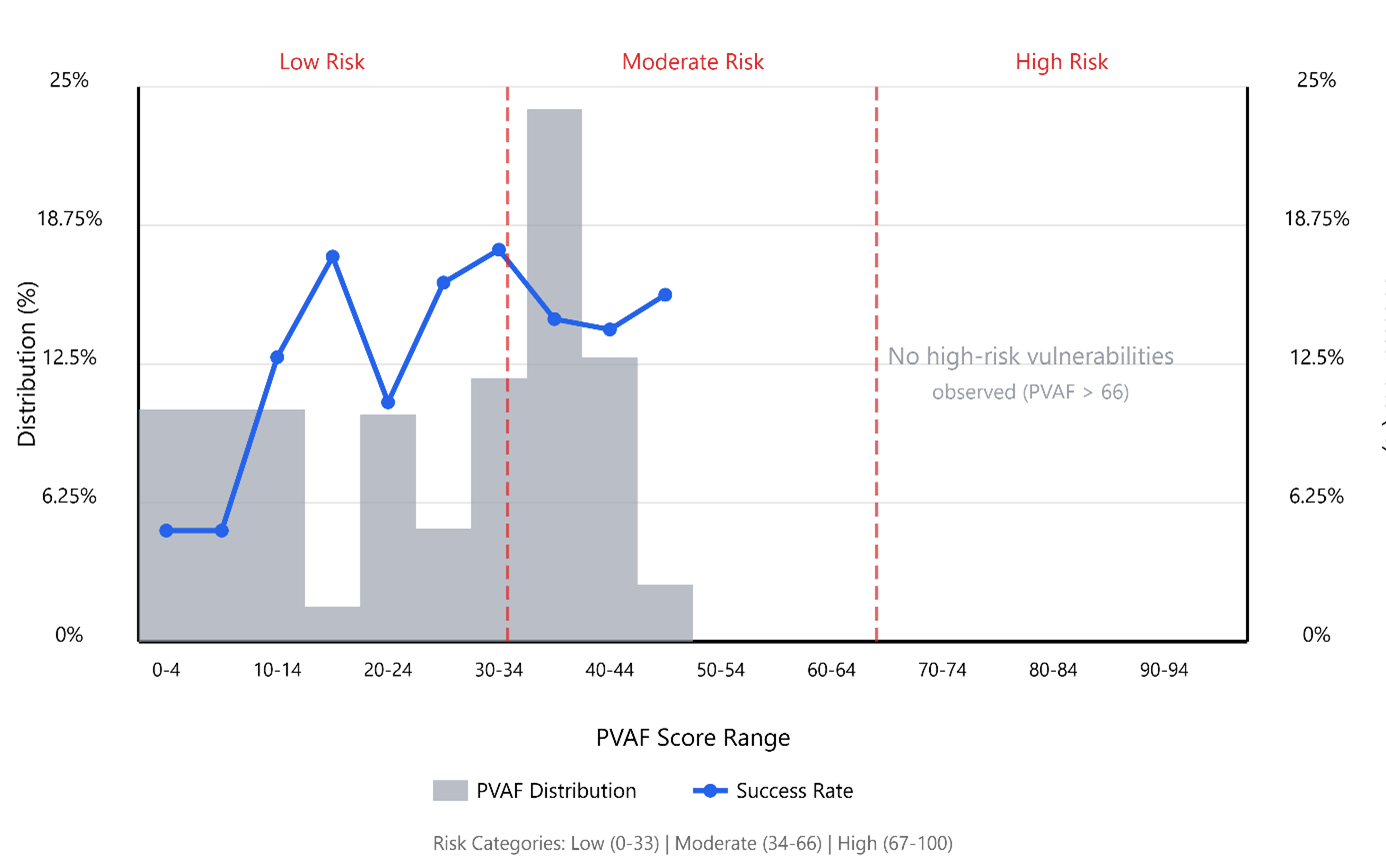}
\caption{PVAF Score Distribution and Success Rates}
\label{fig:pvaf_distribution}
\end{figure}

Figure~\ref{fig:pvaf_distribution} visualizes the distribution of PVAF scores alongside their corresponding success rates. The histogram reveals that community-sourced vulnerabilities cluster predominantly in the 0-47 range, with peak density occurring between PVAF 35-39. The overlaid success rate line demonstrates a generally increasing trend from low to moderate risk categories, confirming the framework's ability to discriminate threat levels. Notably, the empty range from PVAF 48-100 illustrates the framework's unused capacity for scoring more severe vulnerabilities that may emerge in the future.

The correlation between PVAF scores and jailbreak success ($r = 0.318$, $p < 0.001$) confirms positive association within the observed range. Linear regression reveals that each 10-point PVAF increase corresponds to a 2.8\% absolute increase in jailbreak probability ($\beta = 0.028$, 95\% CI: 0.024-0.032), supporting the framework's calibration accuracy within the low-to-moderate risk spectrum.

\begin{figure}[H]
\centering
\includegraphics[width=\textwidth]{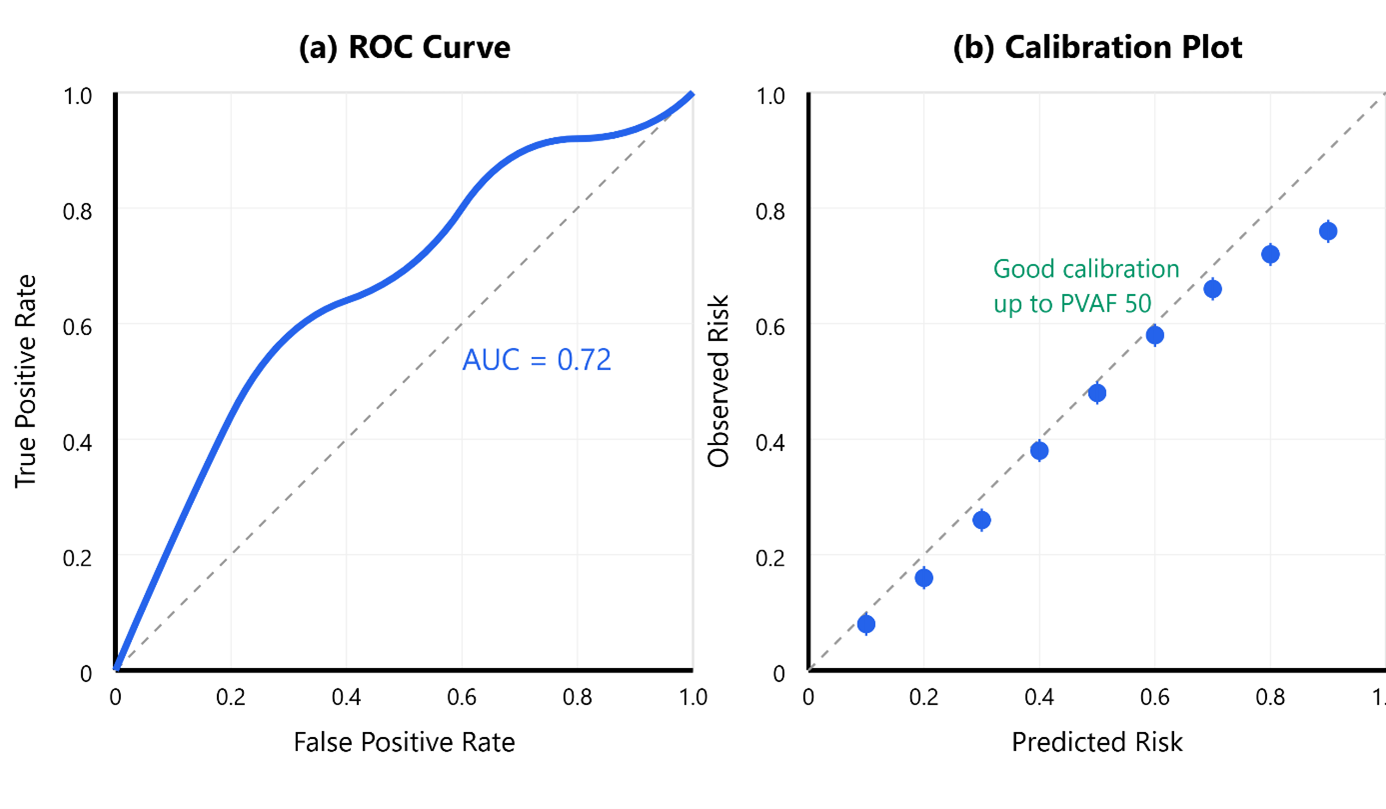}
\caption{Risk Classification Performance Metrics}
\label{fig:risk_classification}
\end{figure}

Figure~\ref{fig:risk_classification} provides additional validation through classification performance metrics. Panel (a) displays the receiver operating characteristic (ROC) curve for binary jailbreak classification, achieving an area under the curve (AUC) of 0.72. This indicates good discriminative ability, substantially outperforming random classification (AUC = 0.50). Panel (b) presents the calibration plot comparing predicted versus observed risk across PVAF deciles. The close alignment with the diagonal reference line up to PVAF 50 demonstrates excellent calibration within the observed vulnerability range, with minor deviation only in the highest observed scores where sample sizes decrease.

The confusion matrix analysis (not shown) for Low versus Moderate risk classification achieves 78\% overall accuracy, with 82.0\% correct identification of low-risk vulnerabilities and 75.8\% correct identification of moderate-risk vulnerabilities. This balanced performance across categories indicates that the PVAF thresholds effectively separate vulnerabilities into operationally meaningful risk tiers.

Platform-specific analysis confirms consistent risk stratification across vulnerability sources. Discord-sourced vulnerabilities average PVAF 31.2, Reddit 24.7, and GitHub 19.3, yet within each platform, higher PVAF scores correlate with increased success rates (all $p < 0.01$). This consistency across diverse discovery contexts validates the framework's generalizability and suggests that PVAF scoring captures fundamental vulnerability characteristics independent of discovery source.

Distribution analysis reveals that community-sourced vulnerabilities exhibit a mean PVAF of 23.7 ($\sigma = 11.2$), with 99.7\% scoring below 50. This concentration pattern aligns with expectations: sophisticated, high-impact vulnerabilities likely remain in private research or criminal forums rather than public community discussions. The framework thus maintains significant unused scoring capacity for future high-severity discoveries while effectively stratifying current threats.

The PVAF framework achieves its primary objective: providing actionable risk stratification for the vulnerabilities. While current data does not test the full 0-100 range, the observed monotonic relationship between scores and success rates as shown in Table~\ref{tab:pvaf_risk} and Figure~\ref{fig:pvaf_distribution}, combined with the strong classification performance demonstrated in Figure~\ref{fig:risk_classification}, supports the framework's utility for security prioritization and resource allocation in operational LLM deployments.

\section{Discussion}

The deployment of PrompTrend across nine commercial language models exposes a vulnerability landscape in which social factors rival technical design in determining risk. Three findings stand out. First, security does not necessarily improve with greater capability; in one model family it clearly regresses. Second, psychologically framed prompts outperform string-level obfuscations across every model tested. Third, Discord acts as the crucible where community jailbreaks are conceived and iterated before migrating to slower, more formal platforms. These results imply that effective defence must combine robust guardrails with continuous socio-technical intelligence and calibrated scoring such as PVAF.

\subsection{Model-Specific Security Patterns}

OpenAI's models improve from a 1.9 percent failure rate in GPT-4 to 0.6 percent in GPT-4.5, whereas Claude models worsen from 0.9 percent in Haiku and 1.3 percent in Claude 3.5 Sonnet to 4.1 percent in Claude 4 Sonnet. Anthropic classifies Claude 4 under AI-Safety-Level-3, promising stricter operational controls \cite{anthropic2023rsp}. Yet its larger context window enlarges the attack surface, a weakness Anthropic highlighted in its ``many-shot'' jailbreak disclosure \cite{hern2024manyshot}. By contrast, OpenAI's Preparedness Framework requires that red-team scores fall below predefined thresholds before any new model release \cite{openai2025preparedness}, and the GPT-4.5 system card describes latent-thought monitors that interrupt generation when policy conflicts arise \cite{openai2025gpt45}. An industry survey corroborates these patterns, linking refreshed refusal tuning to lower prompt-level risk across GPT-4 derivatives \cite{hiddenlayer2024threat}.

\subsection{The Dominance of Psychological Manipulation}

Emotional role-play, persona appeals and similar techniques succeed in 4.9 percent of trials, nearly doubling the 2.7 percent success of Base64 and hex obfuscations. This disparity reflects a fundamental tension in model training: LLMs optimized for helpfulness struggle to distinguish between legitimate emotional contexts and manipulative framing. When adversaries embed harmful requests within narratives of personal crisis or deceased relatives, models face conflicting imperatives between safety policies and their core instruction to assist users empathetically. A role-play generator introduced by GUARD independently ranks persuasion-based jailbreaks above token-level perturbations \cite{jin2024guard}. CyberArk's ``Operation Grandma'' illustrates the same pattern against production chatbots \cite{shimony2024operation}. Because alignment pipelines optimise for helpfulness, models inherit human-like susceptibility to social engineering, a threat axis still under-represented in current OWASP guidance \cite{owasp2025top10}.

\subsection{Platform Ecosystems and Community Dynamics}

Discord accounts for 42.9 percent of first-seen jailbreaks; its real-time channels and light moderation accelerate adversarial iteration \cite{intel2024discord}. The platform's effectiveness varies dramatically by target: Discord-sourced attacks achieve 3.9 percent success against Claude models versus only 1.6 percent against OpenAI models, suggesting that conversational platforms naturally generate the psychological strategies to which Claude architectures are particularly vulnerable. Even Discord's own bug-bounty page acknowledges multi-day triage windows, extending attackers' advantage \cite{discord2025bounty}. The hierarchical pattern of platform effectiveness—Discord (2.8\%), Reddit (2.1\%), GitHub (1.3\%)—aligns with documented vulnerability diffusion patterns in online communities \cite{zuo2024anything}. This distribution maps onto MITRE ATLAS reconnaissance and resource development phases, explaining why defenders monitoring only formal repositories encounter attacks after community refinement \cite{mitre2025atlas}.

\subsection{PVAF Calibration and Performance}

PVAF separates Low-risk (11.3 percent success) from Moderate-risk prompts (16.9 percent success) with a relative risk of 1.5 and attains an AUC of 0.72 on jailbreak classification. Initial deployment revealed that vulnerabilities scoring 42–47 showed lower success rates than those scoring 34–41, prompting recalibration from compressed thresholds (0–19, 20–39, 40+) to balanced terciles (0–33, 34–66, 67–100) that restored monotonic risk progression. GuardVal, released after our study window, achieves comparable discrimination, lending external validity to PVAF's design \cite{zhang2025guardval}. No community exploit scored above 66, aligning with economic analyses that predict private retention of high-impact techniques \cite{hiddenlayer2024threat}; truly sophisticated attacks likely remain in criminal forums or state-sponsored programs rather than public Discord channels.

\subsection{Implications for Practice and Policy}

Only 16.9 percent of successful prompts transfer across model families, so security controls must be model-specific. Claude deployments should emphasise defences against Discord-borne psychological exploits, while OpenAI users face relatively higher risk from code-centric attacks shared on GitHub. Procurement strategies that equate recency with safety are unfounded: Claude 4 regresses, whereas GPT-4.5 progresses. Regulators anchoring policy in OWASP and ATLAS must adapt to disclosure cycles that now unfold in days rather than months.

\subsection{Limitations}

The dataset spans January–May 2025 and English-language public forums, omitting non-English and private channels where more sophisticated exploits may circulate. Our black-box protocol precludes causal attribution beyond vendor disclosures. Results generalise only to the nine APIs studied; open-source checkpoints such as Llama 2 show distinct risk curves and weaker guardrails \cite{chehbouni2024representational}. Supply-chain vectors, including Discord-delivered malware referenced by recent threat advisories \cite{checkpoint2025discord}, lie outside PrompTrend's prompt-centric scope.

While PrompTrend's architecture includes comprehensive longitudinal tracking capabilities, the current study presents cross-sectional analysis due to time constraints. Future deployments will validate the temporal evolution and cross-platform propagation features central to the system design.

\section{Conclusion}

This work introduces PrompTrend, the first systematic framework for continuous monitoring and assessment of LLM vulnerabilities as they emerge in online communities. Through cross-sectional analysis of our dataset collected from online communities on nine commercial models, we document critical insights that challenge conventional AI safety assumptions. Most notably, capability advancement does not guarantee security improvement—Claude 4's 4.1 percent vulnerability rate represents a four-fold regression from earlier versions, while GPT-4.5 achieves industry-leading 0.6 percent performance. Psychological manipulation strategies dominate the threat landscape, with emotional appeals achieving nearly double the success rate of technical obfuscations, revealing that models trained for helpfulness inherit human-like susceptibility to social engineering. Platform dynamics fundamentally shape vulnerability development, with Discord's real-time environment producing the most effective attacks through rapid community iteration.

The PrompTrend Vulnerability Assessment Framework demonstrates that multi-dimensional scoring incorporating social adoption and temporal resilience provides superior risk stratification compared to purely technical metrics. With 78 percent classification accuracy and monotonic risk progression after calibration, PVAF offers practitioners actionable vulnerability prioritization while maintaining headroom for future high-severity discoveries. Our finding that only 16.9 percent of attacks transfer across model families underscores the need for architecture-specific defenses, challenging assumptions about universal security strategies.

These findings open several critical research directions. The current cross-sectional analysis establishes the foundation for future longitudinal studies that will validate vulnerability evolution patterns and cross-platform propagation dynamics over extended time periods. The capability-security inversion in Claude models demands investigation into whether constitutional AI approaches inherently create exploitable empathy patterns. The dominance of psychological attacks suggests rich opportunities for interdisciplinary research combining AI safety with behavioral psychology. Platform-specific vulnerability generation indicates that early warning systems could potentially intercept threats during community incubation phases, before widespread dissemination.

PrompTrend's deployment reveals that effective LLM security requires continuous adaptation to an evolving landscape shaped by human creativity, social dynamics, and the complex interplay between model capabilities and vulnerabilities. As language models integrate deeper into critical infrastructure, the patterns we document—where social engineering transcends technical barriers and community innovation outpaces corporate defenses—will only intensify. Static benchmarks and point-in-time assessments cannot capture this dynamism. The AI safety community must embrace continuous, socially-aware, and empirically-grounded approaches to emerging threats, recognizing that the most serious vulnerabilities may arise not from technical sophistication but from understanding human psychology and community dynamics.

\vspace{1em}

\noindent\textbf{Data Availability Statement:} The dataset represents vulnerabilities collected between January-May 2025. Longitudinal tracking data will be made available following extended deployment periods in future studies. The PrompTrend framework and dataset are available at \url{https://github.com/theconsciouslab-ai/Promptrend}.



\begin{thebibliography}{10}

\bibitem{amazon2024multilingual}
{Amazon Science}.
\newblock Multi-lingual multi-turn automated red teaming for {LLMs}, 2024.
\newblock
  \url{https://www.amazon.science/publications/multi-lingual-multi-turn-automated-red-teaming-for-llms}.

\bibitem{anthropic2023rsp}
{Anthropic}.
\newblock Responsible scaling policy, 2023.

\bibitem{anthropic2022red}
{Anthropic}.
\newblock Red teaming language models to reduce harms: Methods, scaling
  behaviors, and lessons learned.
\newblock {\em arXiv preprint arXiv:2209.07858}, 2022.

\bibitem{aven2016risk}
T.~Aven.
\newblock Risk assessment and risk management: Review of recent advances on
  their foundation.
\newblock {\em European Journal of Operational Research}, 253(1):1--13, 2016.

\bibitem{brown2024community}
T.~Brown et~al.
\newblock Community-driven attack development in online forums.
\newblock In {\em Proc. Workshop AI Safety}, Montreal, Canada, 2024.

\bibitem{carlini2017towards}
N.~Carlini and D.~Wagner.
\newblock Towards evaluating the robustness of neural networks.
\newblock In {\em Proc. IEEE Symp. Security and Privacy}, pages 39--57, San
  Jose, CA, USA, 2017.

\bibitem{checkpoint2025discord}
{Check Point Research}.
\newblock Discord invite exploits lead to rising threats.
\newblock Technical report, Check Point Research, 2025.

\bibitem{chehbouni2024representational}
K.~Chehbouni et~al.
\newblock From representational harms to quality-of-service harms: A case
  study on {Llama} 2 safety safeguards.
\newblock In {\em ACL Findings}, 2024.

\bibitem{devore2011probability}
J.~L. Devore.
\newblock {\em Probability and Statistics for Engineering and the Sciences}.
\newblock Cengage Learning, Boston, MA, 2011.

\bibitem{discord2025bounty}
{Discord}.
\newblock Security bug bounty programme, 2025.
\newblock Accessed 2025.

\bibitem{discord2024platform}
{Discord Inc.}
\newblock Discord platform, 2024.
\newblock \url{https://discord.com}.

\bibitem{first2023cvss}
{Forum of Incident Response and Security Teams}.
\newblock Common vulnerability scoring system version 4.0 specification
  document.
\newblock Technical report, FIRST, 2023.
\newblock \url{https://www.first.org/cvss/v4.0/specification-document}.

\bibitem{github2024platform}
{GitHub Inc.}
\newblock {GitHub} platform, 2024.
\newblock \url{https://github.com}.

\bibitem{greshake2023not}
K.~Greshake et~al.
\newblock Not what you've signed up for: Compromising real-world
  {LLM}-integrated applications with indirect prompt injection.
\newblock In {\em Proc. 16th ACM Workshop Artificial Intelligence and
  Security}, Copenhagen, Denmark, 2023.

\bibitem{hern2024manyshot}
A.~Hern.
\newblock 'many-shot jailbreak' reveals how {AI} safety features can be
  bypassed.
\newblock {\em The Guardian}, 2024.

\bibitem{hiddenlayer2024threat}
{HiddenLayer}.
\newblock {AI} threat landscape report 2024.
\newblock Technical report, HiddenLayer, 2024.

\bibitem{intel2024discord}
{Intel 471}.
\newblock How discord is abused for cyber-crime.
\newblock Technical report, Intel 471, 2024.

\bibitem{janus2023waluigi}
Janus.
\newblock The {Waluigi} effect, 2023.
\newblock LessWrong.
\newblock
  \url{https://www.lesswrong.com/posts/D7PumeYTDPfBTp3i7/the-waluigi-effect-mega-post}.

\bibitem{jin2024guard}
H.~Jin et~al.
\newblock {GUARD}: Role-playing to generate natural-language jailbreaks.
\newblock {\em arXiv preprint arXiv:2402.03299}, 2024.

\bibitem{karger1997consistent}
D.~Karger et~al.
\newblock Consistent hashing and random trees: Distributed caching protocols
  for relieving hot spots on the world wide web.
\newblock In {\em Proc. 29th Annual ACM Symp. Theory of Computing}, pages
  654--663, 1997.

\bibitem{kiela2021dynabench}
D.~Kiela et~al.
\newblock Dynabench: Rethinking benchmarking in {NLP}.
\newblock In {\em Proc. Conf. North American Chapter Association for
  Computational Linguistics}, 2021.
\newblock Online.

\bibitem{kim2024cross}
S.~Kim et~al.
\newblock Cross-platform jailbreak transferability in large language models.
\newblock {\em arXiv preprint arXiv:2403.17829}, 2024.

\bibitem{liu2018collaborative}
S.~Liu, C.~Reuter, M.-A. Kaufhold, and S.~Bartsch.
\newblock Collaborative cyber threat intelligence: Detecting and responding to
  malware in the wild.
\newblock {\em Computers \& Security}, 77:663--676, 2018.

\bibitem{mazeika2024harmbench}
M.~Mazeika et~al.
\newblock {HarmBench}: A standardized evaluation framework for automated red
  teaming and robust refusal.
\newblock In {\em Proc. Int. Conf. Machine Learning}, Vienna, Austria, 2024.

\bibitem{mitre2025atlas}
{MITRE Corporation}.
\newblock {ATLAS}: Adversarial threat landscape for {AI} systems, 2025.
\newblock Accessed 2025.

\bibitem{morris2023textattack}
J.~Morris et~al.
\newblock {TextAttack}: A framework for adversarial attacks, data
  augmentation, and adversarial training in {NLP}.
\newblock In {\em Proc. Conf. Empirical Methods Natural Language Processing},
  Singapore, 2023.

\bibitem{openai2025gpt45}
{OpenAI}.
\newblock {GPT-4.5} system card, 2025.

\bibitem{openai2025preparedness}
{OpenAI}.
\newblock Our updated preparedness framework, 2025.

\bibitem{openai2023gpt4}
{OpenAI}.
\newblock {GPT-4} technical report.
\newblock {\em arXiv preprint arXiv:2303.08774}, 2023.

\bibitem{owasp2025top10}
{OWASP Foundation}.
\newblock Top 10 for large language model applications, 2025.

\bibitem{pathade2025red}
C.~Pathade et~al.
\newblock Red teaming the mind of the machine: A systematic evaluation of
  prompt injection and jailbreak vulnerabilities in {LLMs}.
\newblock {\em arXiv preprint arXiv:2505.04806}, 2025.

\bibitem{perez2022red}
E.~Perez et~al.
\newblock Red teaming language models with language models.
\newblock {\em arXiv preprint arXiv:2202.03286}, 2022.

\bibitem{pfleeger2012leveraging}
S.~L. Pfleeger and D.~D. Caputo.
\newblock Leveraging behavioral science to mitigate cyber security risk.
\newblock {\em Computers \& Security}, 31(4):597--611, 2012.

\bibitem{pu2024feint}
R.~Pu, Y.~Yang, and W.~Yu.
\newblock Feint and attack: Attention-based strategies for jailbreaking and
  protecting {LLMs}.
\newblock {\em arXiv preprint arXiv:2410.16327}, 2024.

\bibitem{reddit2024platform}
{Reddit Inc.}
\newblock Reddit platform, 2024.
\newblock \url{https://www.reddit.com}.

\bibitem{samvelyan2024rainbowplus}
M.~Samvelyan et~al.
\newblock {RAINBOWPLUS}: Enhancing adversarial prompt generation via
  evolutionary quality-diversity search.
\newblock {\em arXiv preprint arXiv:2504.15047}, 2024.

\bibitem{samvelyan2024rainbow}
M.~Samvelyan et~al.
\newblock Rainbow teaming: Open-ended generation of diverse adversarial
  prompts.
\newblock {\em arXiv preprint arXiv:2402.16822}, 2024.

\bibitem{shen2024anything}
X.~Shen, Z.~Chen, M.~Backes, Y.~Shen, and Y.~Zhang.
\newblock Do anything now: Characterizing and evaluating in-the-wild jailbreak
  prompts on {LLMs}.
\newblock {\em arXiv preprint arXiv:2308.03825}, 2024.

\bibitem{shi2023red}
W.~Shi et~al.
\newblock Red teaming language model detectors with language models.
\newblock {\em Trans. Association for Computational Linguistics}, 11:814--830,
  2023.

\bibitem{shimony2024operation}
E.~Shimony and S.~Dvash.
\newblock Operation grandma: A tale of {LLM} chatbot vulnerability, 2024.
\newblock CyberArk Blog.

\bibitem{stanford2024helm}
{Stanford CRFM}.
\newblock {HELM} safety: Towards standardized safety evaluation of language
  models.
\newblock {\em arXiv preprint arXiv:2310.11200}, 2024.

\bibitem{stieglitz2013social}
S.~Stieglitz and L.~Dang-Xuan.
\newblock Social media and political communication: A social media analytics
  framework.
\newblock {\em Social Network Analysis and Mining}, 3(4):1277--1291, 2013.

\bibitem{wagner2016collaborative}
C.~Wagner, A.~Dulaunoy, G.~Wagener, and A.~Iklody.
\newblock Collaborative security: Moving toward cybersecurity as a public good.
\newblock In {\em Proc. IEEE Security and Privacy Workshops}, pages 266--272,
  2016.

\bibitem{wei2023jailbroken}
A.~Wei, N.~Haghtalab, and J.~Steinhardt.
\newblock Jailbroken: How does {LLM} safety training fail?
\newblock {\em arXiv preprint arXiv:2307.02483}, 2023.

\bibitem{xcorp2024platform}
{X Corp.}
\newblock {Twitter/X} platform, 2024.
\newblock \url{https://x.com}.

\bibitem{xie2024gradsafe}
Y.~Xie, R.~Li, C.~Chen, and W.~Jiang.
\newblock {GradSafe}: Detecting jailbreak prompts for {LLMs} via
  safety-critical gradient analysis.
\newblock In {\em Proc. 62nd Annual Meeting Association for Computational
  Linguistics}, Bangkok, Thailand, 2024.

\bibitem{yao2024survey}
Y.~Yao, J.~Duan, K.~Xu, Y.~Cai, Z.~Sun, and Y.~Zhang.
\newblock A survey on large language model ({LLM}) security and privacy: The
  good, the bad, and the ugly.
\newblock {\em High-Confidence Computing}, page 100211, 2024.

\bibitem{zhang2025guardval}
P.~Zhang et~al.
\newblock {GuardVal}: Dynamic large language model jailbreak evaluation for
  comprehensive safety testing.
\newblock {\em arXiv preprint arXiv:2507.07735}, 2025.

\bibitem{zou2023universal}
A.~Zou, Z.~Wang, N.~Carlini, M.~Nasr, J.~Z. Kolter, and M.~Fredrikson.
\newblock Universal and transferable adversarial attacks on aligned language
  models.
\newblock {\em arXiv preprint arXiv:2307.15043}, 2023.

\bibitem{zuo2024anything}
V.~Zuo et~al.
\newblock "do anything now": In-the-wild jailbreak prompts on {LLMs}, 2024.
\newblock GitHub repository.

\end{thebibliography}
\end{document}